\documentclass[9pt,twocolumn,twoside]{osajnl}

\journal{noname} 

\setboolean{shortarticle}{true}

\title{Quantum correlation measurement of laser power noise below shot noise}

\author[1,*]{Jasper R. Venneberg}
\author[1]{Benno Willke}

\affil[1]{Max-Planck-Institut für Gravitationsphysik (Albert-Einstein-Institut) und Institut für Gravitationsphysik, Leibniz Universität Hannover, Callinstraße 38, 30167 Hannover, Germany}

\affil[*]{jasper.venneberg@aei.mpg.de}

\usepackage[normalem]{ulem}

\begin{abstract}
In this Letter, the quantum correlation measurement technique as a method of power noise monitoring is investigated. Its principal idea of correlating two photodetector signals is introduced and contrasted to the conventional approach, which uses only a single photodetector. We discuss how this scheme can be used to obtain power noise information below the shot noise of the detected beam and also below the electronic dark noise of the individual photodetectors, both of which is not possible with the conventional approach. Furthermore, experimental results are presented, that demonstrate a detection of technical laser power noise one order of magnitude below the shot noise of the detected beam.
\end{abstract}

\setboolean{displaycopyright}{false}

\begin{document}

\maketitle

%
Many high precision experiments, such as gravitational wave detectors (GWDs), employ intricate optical readout schemes. These commonly rely on lasers with high power stability, as laser power noise can couple to the readout channel and limit the experiment's sensitivity. Usually, active noise suppression has to be applied and in order to ensure that the stability demand is met, power noise monitoring is necessary. Both require high precision power noise sensing, as the noise of the detector used for stabilization (in-loop sensor) dictates the achievable power stability and the detectors used for monitoring (out-of-loop sensors) must be at least as sensitive as the in-loop sensor to verify the control loop performance. The conventional approach of measuring power noise via a photodetector is fundamentally limited by the shot noise of the detected beam. Thus, power stabilization using conventional photodetection can only achieve a stability equal to the relative shot noise (RSN) of the in-loop beam, which scales proportionally to $P^{-1/2}$ (detected power $P$). For applications such as the aLIGO GWDs \cite{Kwee:12}, which need a power stability of $2\times10^{-9}\, \mathrm{Hz}^{-1/2}$ at $10\, \mathrm{Hz}$, this means a detected power of more than $100\, \mathrm{mW}$, which is beyond the power tolerance of suitable photodetectors. For the current generation of GWDs this is solved by splitting the light onto an array of photodetectors \cite{Kwee:09}, but as the requirements for future generations are likely to be roughly an order of magnitude more stringent, this approach would result in massive technical effort. Therefore, alternative approaches of power noise sensing have been investigated, including carrier attenuation by measuring the power noise in reflection of an optical resonator \cite{Kaufer:19}, injecting squeezed light into the detection port \cite{Vahlbruch_2018} and detecting power noise via radiation pressure \cite{TradNery:21}. While these methods are promising for in-loop detection, they still require considerable effort, and as the out-of-loop detectors have to match the in-loop detector's sensitivity, such an implementation for all detectors would overall result in an enormous technical challenge. In this letter we propose an alternative for out-of-loop detection at a reduced technical effort.

The idea elaborated here, called quantum correlation measurement (QCM), is an approach to significantly reduce the complexity of out-of-loop sensors as less power needs to be detected compared to the in-loop detector. Also, the QCM method is superior in applications where only a limited amount of laser power can be spared for monitoring. It can as well be used in the characterization of bright squeezed light, where standard homodyne detection is limited by local oscillator power noise that beats with the strong carrier of the bright squeezed light field. In the QCM technique, a single photodetector is substituted by a pair of photodetectors, which allows to deduce the power noise from correlating the two detectors' signals. By cancellation of quantum noise contributions in a cross spectral density measurement this enables measurements below the RSN of the detected power, which is impossible with the conventional method. Additionally, the QCM scheme is less sensitive to the electronic dark noise of the detectors, as the individual dark noise contributions are uncorrelated to one another. In this Letter, a comprehensive, quantum mechanical description of the QCM technique is given and its advantages over the conventional concept are discussed. Finally, this theory is confirmed by experimental results from a test experiment, in which we demonstrate a power noise measurement with a sensitivity of one order of magnitude below the RSN of the total detected power.

%
Without loss of generality, the laser light will be described as a plane, quasi-monochromatic, linearly-polarized electromagnetic wave in vacuum and represented by its electric field Heisenberg operator $\hat{E}(t)$  \cite{Danilishin:12}, which can be described in the quadrature picture. There it is decomposed into two orthogonal components: the amplitude quadrature (\textit{c} subscript; 'cosine quadrature') and the phase quadrature (\textit{s} subscript; 'sine quadrature'):
\begin{equation}
\hat{E}(t) = E_0 [(A_\mathrm{c} + \hat{a}_\mathrm{c}(t)) \cos{\omega_0 t} + (A_\mathrm{s} + \hat{a}_\mathrm{s}(t)) \sin{\omega_0 t}].
\label{eq:E_heisenberg}
\end{equation}
$E_0 = \sqrt{\frac{\hbar \omega_0}{\mathcal{A} c \epsilon_0}}$ represents the electric field amplitude where $\mathcal{A}$ is the effective beam cross-section and $\omega_0$ is the optical carrier frequency. $A_\mathrm{c}$ and $A_\mathrm{s}$ denote the stationary components of each respective quadrature, while $\hat{a}_\mathrm{c}(t)$ and $\hat{a}_\mathrm{s}(t)$ are the variations with time. The sine and cosine function together can be interpreted as the base vectors of a two-dimensional coordinate system (optical phase space), where the amplitude and phase quadrature form the two axes.
This allows for a convenient description, as most optical devices linearly transform the quadrature components, and a superposition of fields is represented as the sum of vectors in optical phase space. Hence, in this Letter, light fields will be represented by their quadrature vectors, e.g. field $\hat{\textbf{\textit{a}}} := \left(\hat{a}_\mathrm{c}, \hat{a}_\mathrm{s}\right)$.

The laser power can be defined as the mean energy per time interval carried by the electric field, which can be derived from an integral of the time-averaged Poynting vector \cite{griffiths_2017} over the effective beam cross-section:
\begin{equation}
\hat{P}(t) = \int_\mathcal{A} c \epsilon_0 \overline{|\hat{E}|^2} \, dA,
\label{eq:poynting}
\end{equation}
where the bar represents the time average over the optical period. As is conventional, the carrier field will be defined as solely in the amplitude quadrature, so $A_\mathrm{s} = 0$. Furthermore, it is assumed that the fluctuations with time are small compared to the carrier amplitude, i.e., $\langle a_\mathrm{c}(t)^2\rangle, \langle a_\mathrm{s}(t)^2\rangle \ll A_\mathrm{c}^2$. This yields:
\begin{equation}
\begin{aligned}
\hat{P}(t) &= \frac{\hbar \omega_0}{2} \left((A_\mathrm{c} + \hat{a}_\mathrm{c}(t))^2 + \hat{a}_\mathrm{s}(t)^2\right), \\
&\approx \frac{\hbar \omega_0}{2} \left(A_\mathrm{c}^2 + 2 A_\mathrm{c} \hat{a}_\mathrm{c}(t)\right).
\end{aligned}
\label{eq:power}
\end{equation}
So the mean power is the stationary term $P_0 := \frac{\hbar \omega_0}{2} A_\mathrm{c}^2$ and $\delta \hat{P}(t) := \hbar \omega_0 A_\mathrm{c} \hat{a}_\mathrm{c}(t)$ represents changes with time. For the following, a description in frequency domain is practical, which can be obtained via a Fourier transform. This results in:
\begin{equation}
\delta \hat{P}(\Omega) := \sqrt{2 \hbar \omega_0 P_0} \, \hat{a}_\mathrm{c}(\Omega),
\label{eq:delta_power_freq}
\end{equation}
where $\hat{a}_\mathrm{c}(\Omega)$ is the two-photon amplitude quadrature operator in frequency domain \cite{Danilishin:12}. As photodiodes generate a photocurrent proportional to their incident power, this shows that the description of the respective amplitude quadrature is equivalent to the photodetector signal. In order to quantify fluctuations, the frequency domain equivalent to the variance, the (single-sided) cross power spectral density (CSD) $S^{\hat{\textbf{\textit{p}}}, \hat{\textbf{\textit{q}}}}_{i, j}(\Omega)$ of two quadratures $\hat{p}_i$ and $\hat{q}_j$, $i, j = \mathrm{c}, \mathrm{s}$, can be defined as \cite{Danilishin:12}: 
\begin{equation}
\pi \delta(\Omega + \Omega') S^{\hat{\textbf{\textit{p}}}, \hat{\textbf{\textit{q}}}}_{i, j}(\Omega) = \frac{1}{2} \langle\hat{p}_i(\Omega) \hat{q}_j(\Omega') + \hat{q}_j(\Omega') \hat{p}_i(\Omega)\rangle, 
\label{eq:CPSD}
\end{equation}
where $\delta$ denotes the Dirac distribution. Its magnitude represents the power of the coherent components in the two quadratures $\hat{p}_i$ and $\hat{q}_j$ while its phase describes the relative phase at each given frequency $\Omega$. Put simply, it is a frequency domain measure of how similar or 'correlated' the two signals are. A special case of the CSD is the commonly used auto power spectral density (usually called power spectral density; PSD), in which both quadratures are identical, rendering the phase information obsolete. As such, the PSD describes the magnitude of fluctuations in a single quadrature and can be used to define the relative power noise $\mathrm{RPN}$ of the field $\hat{\textbf{\textit{a}}}$ via Eq. \ref{eq:delta_power_freq} as
\begin{equation}
\mathrm{RPN}(\Omega) = \sqrt{\frac{2 \hbar \omega_0}{P_0} S^{\hat{\textbf{\textit{a}}}}_{\mathrm{cc}}(\Omega)}.
\label{eq:RPN}
\end{equation}
\begin{figure}
    \centering
    \includegraphics[width=0.7\linewidth]{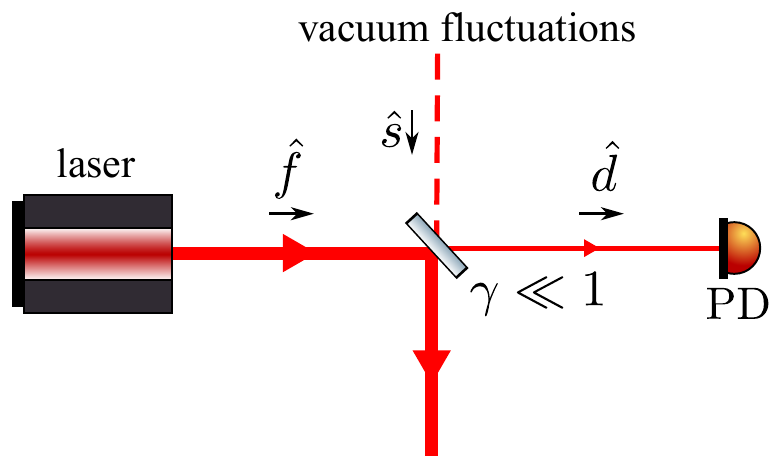}
    \caption{Schematic of a conventional RPN measurement comprising a laser pick-off port via a low-transmission beamsplitter and a single photodetector (PD).}
    \label{fig:setup_trad}
\end{figure}

Here, this formalism shall be applied to determine the limitations of the conventional power noise detection scheme and compare it to the QCM technique. Consider Fig. \ref{fig:setup_trad}. A strong laser field $\hat{\textbf{\textit{f}}}$ of power $P_0$ is incident on a lossless, partly transmissive mirror, which will be treated here as a beamsplitter in quantum mechanical context \cite{Leonhardt_2003}. Only a small fraction of the light is transmitted into the detection field $\hat{\textbf{\textit{d}}}$, which is incident on a photodetector. This field has the power $P_d = \gamma^2 P_0$, where $\gamma$ denotes the amplitude transmission coefficient of the beamsplitter. $\gamma$ is chosen to be small due to the detector's limited power tolerance and to not waste much power for monitoring purposes. The second input of the beamsplitter is an open port and therefore a coupling path for the so called vacuum fluctuations, which represent the minimum excitation of the quantum mechanical electromagnetic field. These random changes of amplitude and phase are present with a finite size even in light states with no coherent amplitude (vacuum states) and are treated here as the vacuum field $\hat{\textbf{\textit{s}}}$. Since $\gamma$ is small, $\hat{\textbf{\textit{s}}}$ is mostly reflected into $\hat{\textbf{\textit{d}}}$. Together with the transmitted field $\hat{\textbf{\textit{f}}}$ one finds
\begin{equation}
\hat{\textbf{\textit{d}}} = \gamma \hat{\textbf{\textit{f}}} + \sqrt{1 - \gamma^2} \hat{\textbf{\textit{s}}}.
\label{eq:traditional_detection_field}
\end{equation}
$\hat{\textbf{\textit{f}}}$ represents the input laser field and thus can contain both technical and quantum noise contributions. $\hat{\textbf{\textit{s}}}$ only carries quantum noise originating from the vacuum fluctuations, which are independent from the carrier field. They appear as a constant contribution to the PSD of the detection field's amplitude quadrature. In this formalism, one finds for a field $\hat{\textbf{\textit{a}}}$ in a vacuum state that $S^{\hat{\textbf{\textit{a}}}}_{\mathrm{cc}}(\Omega) = S^{\hat{\textbf{\textit{a}}}}_{\mathrm{ss}}(\Omega) = 1$ \cite{Danilishin:12}. Thus, one can define the relative shot noise $\mathrm{RSN}$ for a power of $P_0$ in analogy to Eq. \ref{eq:RPN} as:
\begin{equation}
\mathrm{RSN} = \sqrt{\frac{2 \hbar \omega_0}{P_0}},
\label{eq:RSN}
\end{equation}
The RPN of $\hat{\textbf{\textit{d}}}$ can then be calculated using Eqs. \ref{eq:RPN} and \ref{eq:traditional_detection_field}:
\begin{equation}
\begin{aligned}
\mathrm{RPN}_{\hat{\textbf{\textit{d}}}} &= \sqrt{\gamma^2 \frac{2 \hbar \omega_0}{P_d} S^{\hat{\textbf{\textit{f}}}}_{\mathrm{cc}} + \left(1 - \gamma^2\right) \frac{2 \hbar \omega_0}{P_d} S^{\hat{\textbf{\textit{s}}}}_{\mathrm{cc}}}\\
&\approx \sqrt{\mathrm{RPN}_{\hat{\textbf{\textit{f}}}}^2 + \mathrm{RSN}_{P_d}^2} \ge \mathrm{RSN}_{P_d}.
\end{aligned}
\label{eq:traditional_RPN}
\end{equation}
Here, $S_{cc}^{\hat{\textbf{\textit{s}}}} = 1$ and $1 - \gamma^2 \approx 1$ for small beamsplitter transmission was applied. The cross-term containing $S^{\hat{\textbf{\textit{f}}}, \hat{\textbf{\textit{s}}}}_{\mathrm{cc}}$ vanishes, as $\hat{\textbf{\textit{f}}}$ and $\hat{\textbf{\textit{s}}}$ are uncorrelated, since the latter emerges from vacuum fluctuations. This equation shows that $\mathrm{RPN}_{\hat{\textbf{\textit{d}}}}$ is limited by the fluctuations of $\hat{\textbf{\textit{s}}}$, which means that the detection cannot be more sensitive than the RSN of the detected power $P_d$. Eq. \ref{eq:RSN} suggests that this limit can be lowered by increasing $\gamma$ and thus $P_d$, but this can only be done up to the detector's power tolerance. 
%
\begin{figure}
    \centering
    \includegraphics[width=0.8\linewidth]{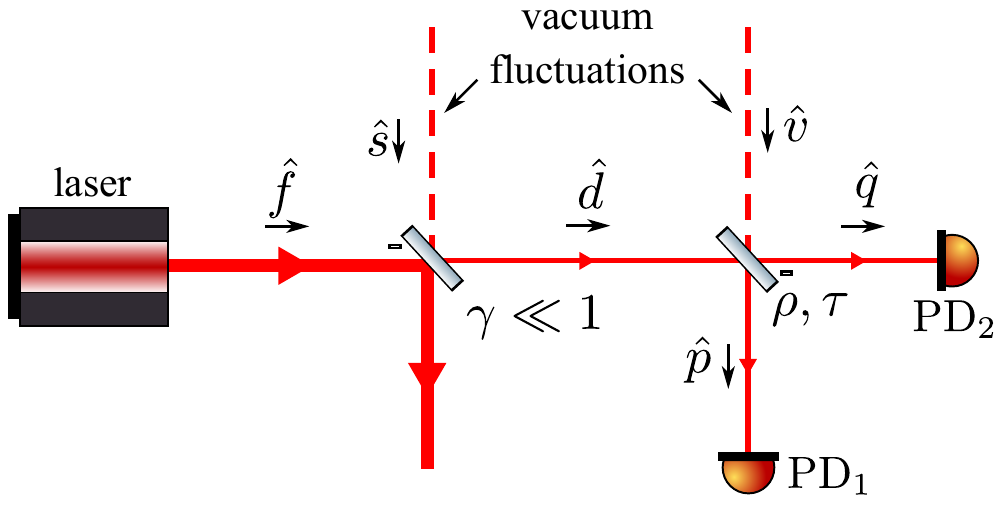}
    \caption{Schematic of the QCM configuration where a beamsplitter and two photodetectors are placed in the detection path.}
    \label{fig:setup_QCM}
\end{figure}

The QCM technique can circumvent this limitation. Similar approaches have been used before, for example for residual noise characterization of the LIGO Livingston detector \cite{martinov17}, measurements of liquid surface fluctuations \cite{mitsui:13}, or spectroscopy of atomic vapors \cite{Mitsui:14}. Here, this method shall be used in a power noise monitoring scheme, which surpasses the shot-noise limit of the conventional scheme for the same detected power. Consider Fig. \ref{fig:setup_QCM}. The first part of this setup is identical to the conventional approach. But instead of sending field $\hat{\textbf{\textit{d}}}$ to one photodetector, the light is split by a beamsplitter (amplitude reflectivity $\rho$ and transmissivity $\tau$) into two beams, that each are incident on a photodetector ($\mathrm{PD}_1$ and $\mathrm{PD}_2$ respectively). In the open port, vacuum fluctuations enter in form of the field $\hat{\textbf{\textit{v}}}$. Together this results in the output fields:
\begin{equation}
\begin{aligned}
\hat{\textbf{\textit{p}}} &= \rho \gamma \hat{\textbf{\textit{f}}} + \rho \sqrt{1 - \gamma^2} \hat{\textbf{\textit{s}}} + \tau \hat{\textbf{\textit{v}}}, \\
\hat{\textbf{\textit{q}}} &= \tau \gamma \hat{\textbf{\textit{f}}} + \tau \sqrt{1 - \gamma^2} \hat{\textbf{\textit{s}}} - \rho \hat{\textbf{\textit{v}}},
\end{aligned}
\label{eq:QCM_fields}
\end{equation}
where the asymmetric beamsplitter convention was applied as denoted in Fig. \ref{fig:setup_QCM}.

This creates the opportunity to measure the CSD of the two photodetectors' signals, which are proportional to the amplitude quadratures of the two fields $\hat{\textbf{\textit{p}}}$ and $\hat{\textbf{\textit{q}}}$ respectively. Eqs. \ref{eq:CPSD} and \ref{eq:QCM_fields} yield
\begin{equation}
S^{\hat{\textbf{\textit{p}}}, \hat{\textbf{\textit{q}}}}_{\mathrm{cc}} = \rho \tau \left( \gamma^2 S^{\hat{\textbf{\textit{f}}}}_{\mathrm{cc}} + (1 - \gamma^2) S^{\hat{\textbf{\textit{s}}}}_{\mathrm{cc}} - S^{\hat{\textbf{\textit{v}}}}_{\mathrm{cc}} \right)
\label{eq:QCM_CSD}
\end{equation}
where again all cross-terms vanish, as $\hat{\textbf{\textit{f}}}$, $\hat{\textbf{\textit{s}}}$ and $\hat{\textbf{\textit{v}}}$ are pairwise uncorrelated. Due to the 180° phase difference between transmitted and reflected fields at the beamsplitter, the vacuum fluctuation terms $S^{\hat{\textbf{\textit{s}}}}_{\mathrm{cc}}$ and $S^{\hat{\textbf{\textit{v}}}}_{\mathrm{cc}}$ have opposing signs. This effect is independent of the used beamsplitter convention and it shall be particularly highlighted here, as this phase difference is crucial for inferring the technical power noise from the CSD. For this, the normalized cross spectral density (NCSD) is defined in an analogous way to the square of the RPN:
\begin{equation}
\begin{aligned}
\mathrm{NCSD}_{\hat{\textbf{\textit{p}}}, \hat{\textbf{\textit{q}}}} &= \frac{2 \hbar \omega_0}{P_{\mathrm{QCM}}} S^{\hat{\textbf{\textit{p}}}, \hat{\textbf{\textit{q}}}}_{\mathrm{cc}} = \rho \tau \gamma^2 \frac{2 \hbar \omega_0}{P_{\mathrm{QCM}}} \left( S^{\hat{\textbf{\textit{f}}}}_{\mathrm{cc}} - 1 \right) \\
&= \mathrm{RPN}_{\hat{\textbf{\textit{f}}}}^2 - \mathrm{RSN}_{P_0}^2.
\end{aligned}
\label{eq:QCM_NCSD}
\end{equation}
Here, $S^{\hat{\textbf{\textit{s}}}}_{\mathrm{cc}} = S^{\hat{\textbf{\textit{v}}}}_{\mathrm{cc}} = 1$ was used. The effective QCM power was defined as $P_{\mathrm{QCM}} = \rho \tau \gamma^2 P_0 = \rho \sqrt{1 - \rho^2} P_d = \sqrt{P_\mathrm{PD_1} P_\mathrm{PD_2}}$ and can be interpreted as the equivalently detected power for this method. Furthermore, it will be assumed that the field $\hat{f}$ contains technical noise in addition to quantum noise of a coherent state, which has the same quantum noise properties as a vacuum state, i.e., $\mathrm{RPN}_{\hat{\textbf{\textit{f}}}}^2 = \left( \mathrm{RPN}^{\mathrm{tech}}_{\hat{\textbf{\textit{f}}}} \right)^2 + \mathrm{RSN}_{P_0}^2$. This leads to
\begin{equation}
\sqrt{\mathrm{NCSD}_{\hat{\textbf{\textit{p}}}, \hat{\textbf{\textit{q}}}}} = \mathrm{RPN}^{\mathrm{tech}}_{\hat{\textbf{\textit{f}}}}.
\label{eq:QCM_tech}
\end{equation}
This remarkable result can be understood as follows. For a quantum noise limited laser beam $S^{\hat{\textbf{\textit{f}}}}_{\mathrm{cc}}$ corresponds to the noise of a coherent state and is equal to the quantum noise of the vacuum state $S^{\hat{\textbf{\textit{s}}}}_{\mathrm{cc}}$. In this case the sum of the first two terms in Eq. \ref{eq:QCM_CSD} associated to $\hat{f}$ and to $\hat{s}$ add up to the PSD of a vacuum state. This means that $\hat{d}$ has the noise of a coherent state.
Due to the 180° phase difference introduced by the second beamsplitter the quantum noise contributions in the photocurrents of PD1 and PD2 associated to $\hat{d}$ and $\hat{s}$ have 180° phase shift with respect to each other and cancel in the CSD calculation (see Eq. \ref{eq:QCM_CSD}). If the incoming beam also carries technical noise, the quantum noise cancels as described and the only remaining contribution is the technical laser noise of the input field $\hat{\textbf{\textit{f}}}$. It should be mentioned that the individual vacuum fields are fundamentally uncorrelated to each other. Therefore, they do not compensate one another in time domain, but their spectral contributions do. Thus, this method differs conceptually from homodyne detection, in which local oscillator contributions cancel also in time domain. As such, the QCM technique is not applicable to reduce shot noise on an in-loop sensor of a power stabilization loop, however it can ease spectral power noise monitoring, e.g. for out-of-loop sensing, substantially.

In summary, this means that the QCM technique is ideally only sensitive to the technical power noise of the input laser field. As such it is advantageous to the conventional method which is limited by the detected beam's RSN. Furthermore, this alternative can be implemented with minimal technical effort by adding a beamsplitter and a second photodetector. Another benefit can be found considering incoherent noise in the two photodetector signals, like the individual electronic dark noise. As these noise components only add uncorrelated crossterms to Eq. \ref{eq:QCM_CSD}, which average out to zero, they theoretically have no influence on the CSD. In a regular PSD measurement, however, dark noise poses a lower limit of the detection sensitivity.

%
So far in this Letter the QCM technique was analyzed theoretically. To demonstrate it in practice, a test experiment was set up to compare the RPN of a laser beam measured by the conventional scheme (Fig. \ref{fig:setup_trad}) and by the QCM technique (Fig. \ref{fig:setup_QCM}). A Nd:YAG solid-state laser in a non-planar ring oscillator configuration with an operating wavelength of $1064\, \mathrm{nm}$ and a full beam power (representing field $\hat{f}$) of $770\, \mathrm{mW}$ fabricated by the company Coherent was used. In both configurations this power was attenuated at the first beamsplitter to $4\, \mathrm{mW}$ in the detection field $\hat{\textbf{\textit{d}}}$ in order to demonstrate a shot noise limited detection for the conventional scheme. For the QCM method $\hat{\textbf{\textit{d}}}$ further was incident on a 50:50 beamsplitter. This resulted in an effective QCM power of $2\, \mathrm{mW}$, which was sufficient for this setup. In general a higher $P_\mathrm{QCM}$ increases the signal-to-noise ratio, when considering noise sources that couple coherently in both signals (like electronic interference). Therefore the best results should be found for a beamsplitter with equal transmitted and reflected power, as it is used here.
\begin{figure}
\centering
\includegraphics[width=\linewidth]{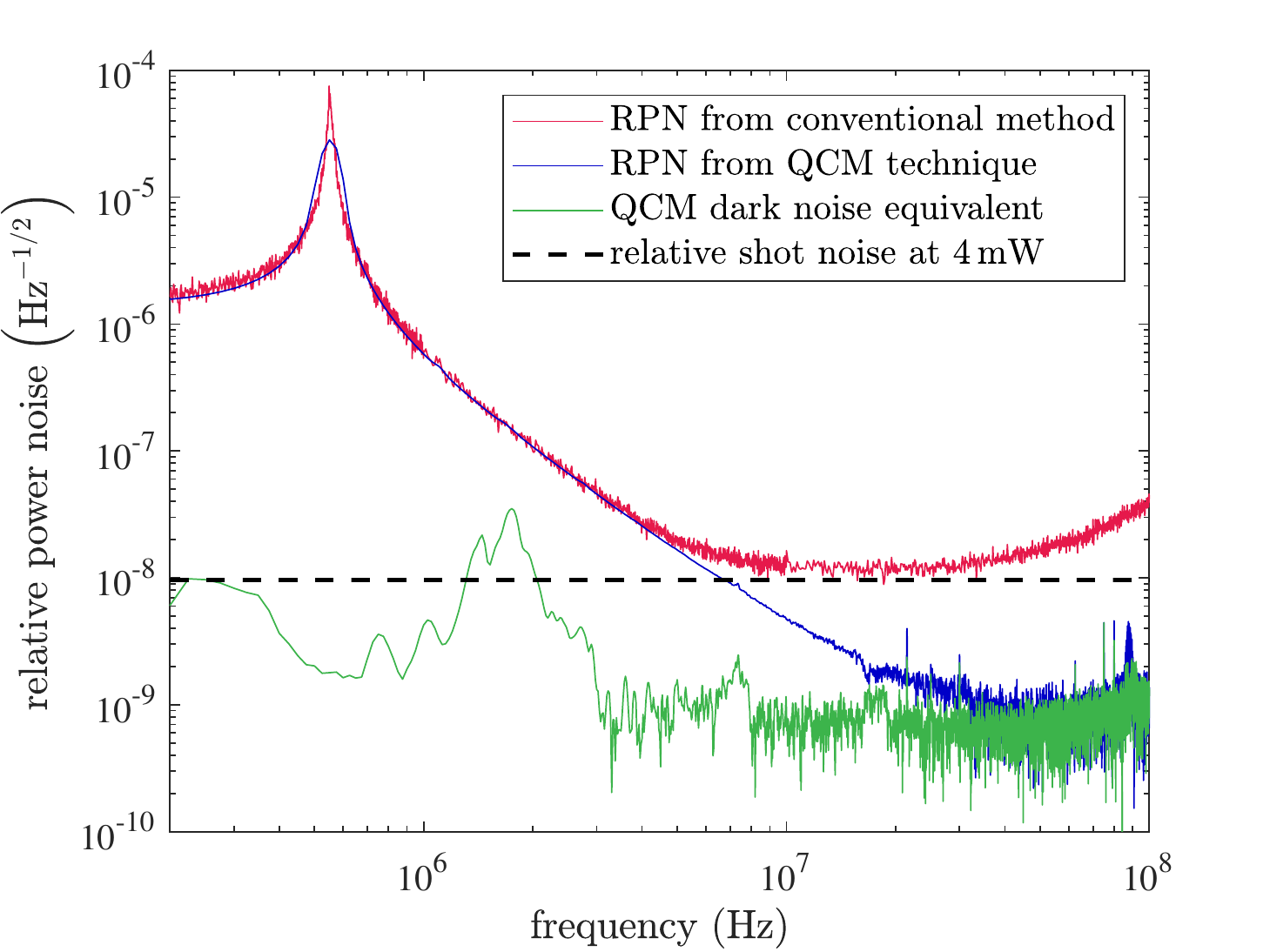}
\caption{Comparison of the RPN of the free running laser noise obtained from a conventional power noise measurement and from the QCM scheme.}
\label{fig:QCMresult}
\end{figure}
With both methods, the free running power noise of the laser was measured in a frequency band from $200\, \mathrm{kHz}$ to $100\, \mathrm{MHz}$. The PSD for the conventional measurement was recorded with a spectrum analyzer. For the QCM method, the analog signals of both photodetectors were digitized and recorded simultaneously with a digital oscilloscope and then processed via the MATLAB function 'cpsd' (which is based on Welch's method) to obtain the CSD. Fig. \ref{fig:QCMresult} illustrates the results of these measurements calibrated to RPN. In red the conventional RPN measurement is depicted. The blue curve shows the square root of the NCSD obtained from the QCM technique, which according to Eq. \ref{eq:QCM_tech} should coincide with the technical power noise of the laser. In the black dashed curve, the RSN of the detection beam at $4\, \mathrm{mW}$ is given. This measurement band was chosen since the technical power noise follows a distinct and well-known behavior dictated by the relaxation oscillation of the laser source, which results in a sharp resonant peak (in this case at roughly $540\, \mathrm{kHz}$) followed by an approximate $1 / f^2$ decline for larger frequencies \cite{siegman86, Kwee:08}. Both the red and blue curve portray this well up to $4\, \mathrm{MHz}$ where they closely agree. The minor apparent difference at the resonance stems from different frequency resolutions and is irrelevant here. 

After the steep decline, the red curve levels on a value of $10^{-8}\, \mathrm{Hz}^{-1/2}$ from about $10\, \mathrm{MHz}$ to $30\, \mathrm{MHz}$. This directly corresponds to the RSN of the detection beam and indicates that the conventional method is shot noise limited in this frequency range. For higher frequencies, the electronic noise of the photodetector becomes larger than the shot noise, which results in the rising noise towards the end of the measurement band. This illustrates the typical limitation of the conventional method both due to shot noise and electronic dark noise as discussed above. 

For the QCM technique, this is different. Like the red curve, the QCM result in blue shows the expected technical laser noise for lower frequencies, but it continues the $1/f^2$ behavior even below the shot noise level of the detection beam, which it meets at around $7\, \mathrm{MHz}$. At roughly $50\, \mathrm{MHz}$ it finally reaches the lowest point at a sensitivity of about $10^{-9}\, \mathrm{Hz}^{-1/2}$, which is one order of magnitude below the RSN. For frequencies above $50\, \mathrm{MHz}$, the QCM curve is limited by the projected dark noise equivalent for this measurement, which is illustrated in green. This dark noise measurement was obtained by performing a measurement without laser power incident on the detectors, while all other circumstances (such as position of the detectors and the measurement electronics, background lighting of the laboratory, etc.) were identical. The term 'equivalent' is used in order to emphasize that this differs from a conventional dark noise measurement (i.e. a PSD for a single photodetector), as this curve is also obtained from an NCSD.

Overall, the QCM technique provides a power noise measurement significantly below the detection beam's RSN. The RPN obtained this way matches the behavior expected from the laser source, which supports the above derived formalism. Furthermore, it indicates the feasibility of the QCM technique as a sub shot noise power noise monitor under realistic circumstances. The limitation of the measurement shown here is the QCM dark noise equivalent. In an infinitely long measurement, this dark noise contribution should tend to zero due to the uncorrelated nature of the sources. However, for realistic measurement times, the contribution to the CSD is finite and expected to scale inversely with the measurement time (due to the derivation of the CSD from Welch's method). With longer measurements this could be reduced to an arbitrarily low level, as long as the corresponding noise in both photodetectors is truly uncorrelated. The setup used here was limited by the processing speed of the available recording and evaluation hardware. A future implementation of a fast Fourier transform algorithm via a field programmable gate array could result in quasi real-time measurements.

%
In this Letter the quantum correlation measurement technique was investigated as an alternative to conventional power noise monitoring, which is both powerful and easily implemented. Its theoretical basis was laid and compared to the conventional concept. Subsequently, a test experiment was described and its results were presented, which support the established theory. In conclusion, the QCM scheme was demonstrated to allow for a sub shot noise measurement of power fluctuations, which could be beneficial for applications with extremely high power stability demands or with limited light power available for monitoring purposes. Another promising application of this technique could be as a bright squeezed light detector, which becomes apparent when considering Eq. \ref{eq:QCM_NCSD} for an input field $\hat{\textbf{\textit{f}}}$ with non-classical noise properties like (anti-) squeezed light.

\noindent\textbf{Funding.} Funded by the Deutsche Forschungsgemeinschaft (DFG, German Research Foundation) under Germany’s Excellence Strategy – EXC-2123 QuantumFrontiers – 390837967.

\noindent\textbf{Acknowledgements.} The authors thank M. Trad Nery for useful comments on this manuscript.

\noindent\textbf{Disclosures.} The authors declare no conflicts of interest.

\bibliography{QCM_paper}

\bibliographyfullrefs{QCM_paper}

\end{document}